\documentclass[newstyle,twocolumn,proceedings]{rmaa}
\newcommand{\ii}{\'{\i}}
\newcommand{\kms}{\,\mbox{km s$^{-1}$}}
\newcommand{\Otwo}{[\ion{O}{2}]~3727\AA}
\newcommand{\Othree}{[\ion{O}{3}]~5007\AA}
\title{Tomography of high-redshift clusters with OSIRIS}
\altaffiltext{1}{Osservatorio Astronomico di Brera, Merate, ITALY.}
\altaffiltext{2}{Marie Curie Fellow}
\altaffiltext{3}{Anglo-Australia Observatory, Epping, AUSTRALIA.}
\altaffiltext{4}{Instituto de F\ii sica de Cantabria, Santander, SPAIN}
\altaffiltext{5}{Universidad de Cantabria, Santander, SPAIN.}
\author{A.~Fern\'andez-Soto\altaffilmark{1,2},
J.~Bland-Hawthorn\altaffilmark{3}, J.~I.~Gonz\'alez-Serrano\altaffilmark{4,5}
and R.~Carballo\altaffilmark{5} }

\fulladdresses{
\item A. Fern\'andez-Soto: OAB, Via Bianchi 46, Merate (LC), 23885-ITALY 
(fsoto@merate.mi.astro.it).
\item J. Bland-Hawthorn: AAO, POBox 296, Epping NSW1710-AUSTRALIA 
(jbh@aaoepp.aao.gov.au).
\item J.I. Gonz\'alez-Serrano: IFCA, Avda. de los Castros s/n, Santander, 
39005-SPAIN (gserrano@ifca.unican.es).
\item R. Carballo: E.T.S.I.C.C.P., Avda. de los Castros s/n , Santander, 
39005-SPAIN (carballor@unican.es). }

\shortauthor{Fern\'andez-Soto et al.}
\shorttitle{Tomography of high-redshift clusters}


\abstract{High-redshift clusters of galaxies are amongst the largest
cosmic structures. Their properties and evolution are key ingredients
to our understanding of cosmology: to study the growth of structure from the
inhomogeneities of the cosmic microwave background; the processes of galaxy
formation, evolution, and differentiation; and to measure the cosmological
parameters (through their interaction with the geometry of the universe, the
age estimates of their component galaxies, or the measurement of the amount
of matter locked in their potential wells). However, not much is yet known
about the properties of clusters at redshifts of cosmological interest. We
propose here a radically new method to study large samples of cluster
galaxies using microslits to perform spectroscopy of huge numbers of objects
in single fields in a narrow spectral range--chosen to fit an emission line
at the cluster redshift. Our objective is to obtain spectroscopy in a very
restricted wavelength range ($\approx$ 100 \AA\ in width) of several
thousands of objects for each single 8x8 square arcmin field. Approximately
100 of them will be identified as cluster emission-line objects and will
yield basic measurements of the dynamics and the star formation in the
cluster (that figure applies to a cluster at $z \approx 0.50$, and becomes
$\approx 40$ and $\approx 20$ for clusters at $z\approx 0.75$ and $z \approx
1.00$ respectively). This is a pioneering approach that, once proven, will be
followed in combination with photometric redshift techniques and applied to
other astrophysical problems. }
\resumen{Los c\'umulos de galaxias a alto redshift est\'an entre las
estructuras c\'osmicas de mayor tama\~no. Sus propiedades y evoluci\'on nos
dan informaci\'on sobre los aspectos claves de la cosmolog\ii a: el
crecimiento de estructura a partir de las semillas del fondo c\'osmico de
microondas, los procesos de formaci\'on, evoluci\'on y diferenciaci\'on de
galaxias, y la medida de los par\'ametros cosmol\'ogicos (a partir de la
interacci\'on de los c\'umulos con la geometr\ii a del universo, la
estimaci\'on de las edades de las galaxias que los forman, o la medida de la
cantidad de materia en sus pozos de potencial). No obstante, no sabemos mucho
sobre las propiedades de los c\'umulos lejanos (en el rango $z \approx
0.5-1.0$). Con esta propuesta presentamos un m\'etodo radicalmente nuevo para
estudiar gran cantidad de galaxias en una muestra significativa de c\'umulos,
mediante espectroscop\ii a en un rango muy restringido (aproximadamente 100
\AA, elegido para contener una l\ii nea de emisi\'on al redshift del
c\'umulo). Combinando el gran tama\~no del campo de OSIRIS con la posibilidad
de utilizar microrrendijas y el peque\~no rango espectral, obtendremos
espectros de miles de objetos por cada campo, de los cuales aproximadamente
100 ser\'an identificados como pertenecientes al c\'umulo (a $z \approx
0.50$; la cifra es $\approx 40$ a $z \approx 0.75$, y $\approx 20$ a $z
\approx 1.00$). Estos objetos nos dar\'an informaci\'on sobre las propiedades
cinem\'aticas y la formaci\'on estelar en el c\'umulo. Este es un m\'etodo
pionero, cuya utilidad, una vez demostrada, puede ser mejorada en
combinaci\'on con medidas fotom\'etricas de redshift, y aplicada a otros
problemas astron\'omicos.}

\listofauthors{A.~Fern\'andez-Soto, J.~Bland-Hawthorn,
J.~I.~Gonz\'alez-Serrano, R.~Carballo}
\indexauthor{Fern\'andez-Soto, A.}
\indexauthor{Bland-Hawthorn, J.}
\indexauthor{Gonz\'alez-Serrao, J.~I.}
\indexauthor{Carballo, R.}

\begin{document}

\maketitle

\section{Introduction}
This contribution introduces a proposal presented to the OSIRIS Scientific 
Committee. The objective of the
proposal is the study of high-redshift clusters ($z \approx 0.5-1.0$)
through an innovative technique that makes use of several features that
render OSIRIS and GTC a unique facility for our purposes:\\
(i) a very large collecting area,\\
(ii) high efficiency in the red spectral region,\\
(iii) a large field of view (8 arcminute side),\\
(iv) availability of tunable filters,\\
(v) availability of nod-and-shuffle/microslits (see Glazebrook and 
Bland-Hawthorn 2001).

The main thrust of our proposal is to obtain moderate-resolution spectroscopy
in a narrow wavelength range (100 \AA) of all objects in the field of view of
OSIRIS. We intend to point the camera to a $z \approx 0.5$ cluster, tune the
wavelength range to observe one of the strong emission lines at the redshift
of the cluster (usually \Otwo) within a velocity range of $\pm 3000 \kms$,
and work in slitless mode to pick up the spectra of all objects in the
field. Our estimates and simulations, described in detail in Section 2,
indicate that in a single three-hour exposure we will pick up approximately
100 emission-line objects pertaining to the cluster. This will allow for a
detailed study of both the kinematic and star-formation properties of the
cluster. The slitless mode also allows us the extra gain of getting all the
light from the objects into the analysis. In Section 3 we briefly introduce
possible changes and/or improvements to be done in a subsequent phase.

\section{Basic description and simulation of the observations}

We have used the OSIRIS simulator, together with IRAF and IDL, to simulate 
the images of a cluster at redshift 0.55 obtained by OSIRIS. The parameters 
of our simulation are as follows:

1- The luminosity function of Coma (L\'opez-Cruz et al 1997), with typical
cluster parameters taken from the literature (Smail et al 1997, Dressler et
al 1999, Fassano et al 2000). No radial segregation of properties is included
in the simulation.

2- Field galaxies are bootstrapped from the catalogue of the Hubble Deep
Field by Fern\'andez-Soto et al (1999), including spectral type, apparent
magnitude, and redshift for each object.

3- For the {\bf direct image}: exposure time of 20 minutes, field
size=8.5x8.5 arcminutes, $V$-band filter, limiting magnitude $V \approx
26.5$.

4- For the {\bf dispersed image}: exposure time of 3 hours, same field size,
narrow-band filter with $\lambda_{\rm c}=5775$ \AA, $\Delta\lambda_{\rm
FWHM}=100$ \AA, grating G1500R (resolution $\Delta v \approx 275 \kms$). The
spectral types and emission line strengths are as given by the simulation,
with the spectral templates in Fern\'andez-Soto et al (1999).

The direct and dispersed images of the field are shown in Figures 1 and
2. Noise has been added to the direct image to mimic the expected detection
limit. Although no noise was added to the dispersed images, all the figures
given hereafter do take into account the expected noise properties of the
images.

\begin{figure*}
  \begin{center}
    \includegraphics[width=10cm]{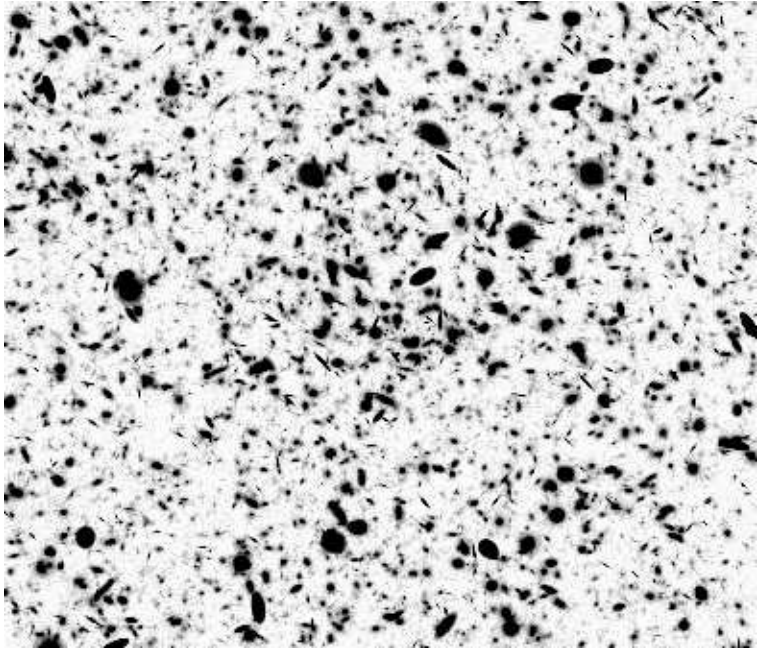}
  \end{center}
  \caption{Direct OSIRIS image of the cluster field.}  
\end{figure*}

\begin{figure*}
  \begin{center}
    \includegraphics[width=10cm]{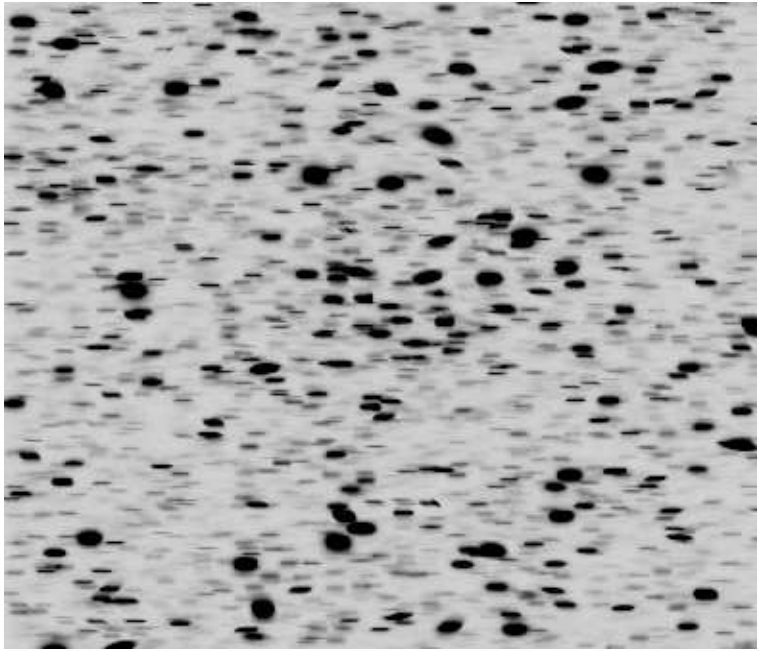}
  \end{center}
  \caption{Dispersed image of the same field.}  
\end{figure*}

Figure 3 shows instead only an enlargement of the central area of the
field. From left to right we show all the galaxies, the field galaxies, and
the cluster galaxies. It is clear from the image that it will be possible to
discern cluster membership based exclusively in the detection of emission
lines. Contamination by other lines will be minimal, with the only possible
interlopers being \Othree\ at much smaller redshift (with a much smaller
volume being probed) and Lyman-$\alpha$ at much higher redshift, which will
render the galaxies dim enough to be undetected in our images.

\begin{figure*}
  \begin{center}
    \includegraphics[width=\textwidth]{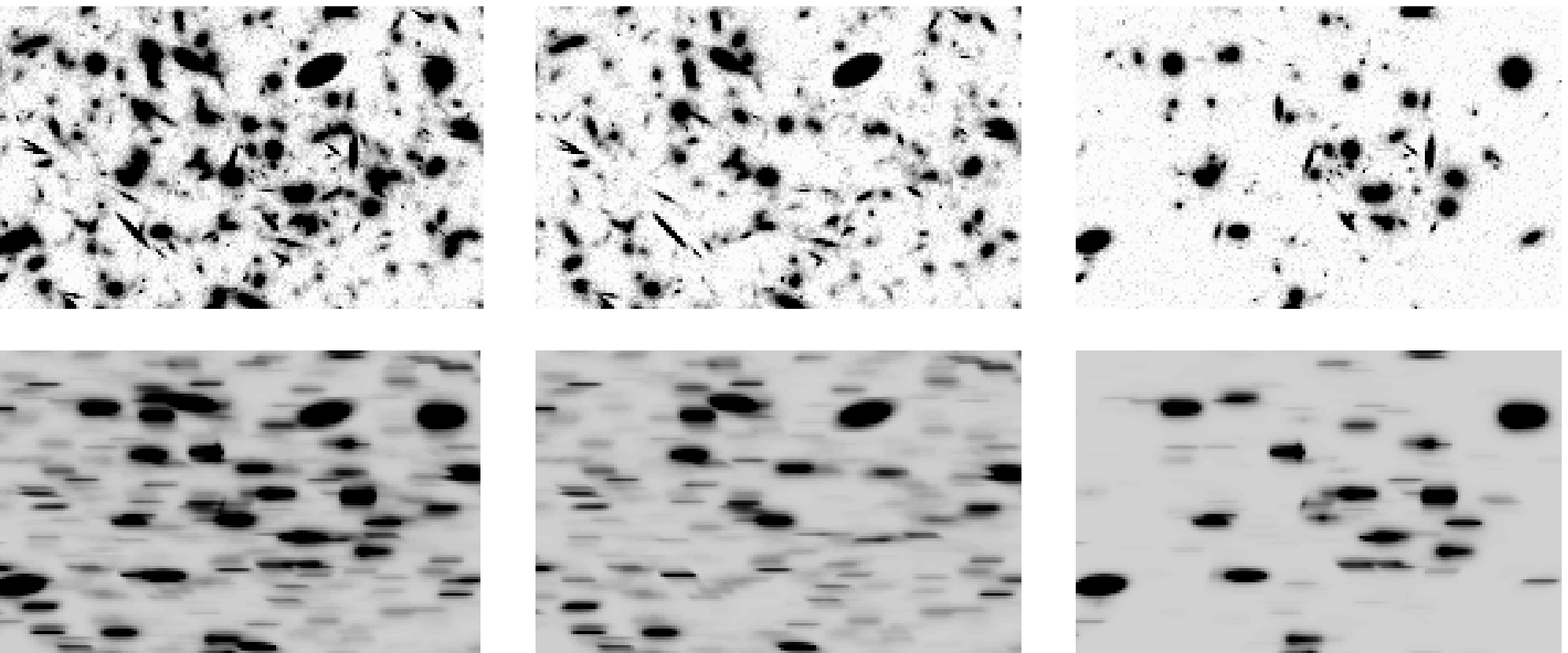}
  \end{center}
  \caption{Expanded view of the central area with all galaxies (left),
the field (center), and the cluster (right).}  
\end{figure*}

Table 1 lists the expected signal-to-noise values with which the continuum
and the emission lines would be detected for the objects in our field, as a
function of both AB magnitude and equivalent width of the emission line.  The
magnitude limits that come out of this calculation for detection (S/N=3 per
resolution element in three hours) in the dispersed image are AB($V$)=25.5
for the continuum, and AB($V$)=25.1, 25.9, 26.6, 27.3 for an emission line
with EW=5, 10, 20, 40 \AA\ respectively.

With these numbers, and using the above-mentioned luminosity functions and EW
distributions, we expect to detect up to $\approx$ 100 objects belonging to
the cluster in a single OSIRIS image centered in the cluster. This number of
objects will represent one of the largest surveys of cluster galaxies ever
observed, and constitute by far the largest study to date of a high-redshift 
cluster.

\begin{table}[!b]
  \begin{center}
    \caption{Signal-to-noise calculations}  
    \begin{tabular}{l r r r r}\hline\hline
AB($V$)          & 24.0 & 25.0 & 26.0 & 27.0 \\ \hline
(S/N)$_{\rm cont}$     &  6.3 &  2.6 &  1.0 &  0.4 \\ 
(S/N)$_{\rm EW=5\AA}$   &  4.3 &  1.8 &  0.7 &  0.3 \\ 
(S/N)$_{\rm EW=10\AA}$  &  8.4 &  3.5 &  1.5 &  0.6 \\ 
(S/N)$_{\rm EW=20\AA}$  & 16.0 &  6.9 &  2.9 &  1.2 \\ 
(S/N)$_{\rm EW=40\AA}$  & 29.0 & 13.0 &  5.6 &  2.3 \\ \hline\hline
    \end{tabular}
  \end{center}
\end{table}

\section{The next steps}

Up to this point we have only presented what we could consider a ``single
building block'' of this project. Several different routes can be followed by
appending ``building blocks'' in different directions. Let us explore some of
them:

1- Move towards higher redshift. This is an obvious direction to move
to. With the exact same settings presented in the previous section, we would
detect $\approx 40$ objects in a $z \approx 0.75$ cluster, and $\approx 20$
in a cluster at $z \approx 1.0$.

2- Move towards larger areas. We can obviously add other fields to the
initial one, hereby mosaicing to obtain a larger field and a complete survey
of the cluster. An added interest of this project is to study the dynamics of
the cluster in the space where its kinematics melts with the Hubble
flow. Theoretical studies (Lilje \& Lahav 1991) show that it is possible to
measure the cosmological parameters using this kind of study.

3- Move to deeper images. This is obviously needed in combination with the
first option above, but can also be useful in the original case if deeper
study of a particular cluster becomes interesting

4- In any move towards higher redshift, we will be entering the red part of
the spectrum, where noise from sky lines becomes dominant. In this case we 
will need to carefully choose the redshift range to study, and use the
tunable filter possibilities to work only in the limited spectral
range available between strong sky lines. This is an excellent means of using
otherwise ``bad'', clear nights.

5- Another improvement will be the use of microslits plus nod-and-shuffle
techniques. It will be useful under two different situations: whenever we
want to use bright or grey nights (when the sky can be almost completely
eliminated by using microslits and sky-free wavelength intervals--otherwise
it will dominate the dispersed image), and/or when we have 
\adjustfinalcols
overcrowded fields
and have some extra information on which objects we prefer to
observe. Glaezebrook and Bland-Hawthorn (2001) show that the gain in sky
darkness obtained in this way amounts to a factor of ten reduction in the
background.

6- A large improvement will come from the combination of this technique with
photometric redshifts. If we can obtain beforehand multi-band images of the
field (say $UBVI$, for example) we can perform a photometric redshift
analysis in order to eliminate those objects whose colours are clearly
incompatible with being at the cluster redshift, and go far deeper by looking
only at the ``interesting'' ones via microslits (as described in 5 above).


\end{document}